\documentclass{moriond}

\def\Journal#1#2#3#4{{#1} {\bf #2}, #3 (#4)}

\def\APJ{\em Astrophys. J.}
\def\MNRAS{\em Mon. Not. R. Astron. Soc.}

\def\be{\begin{equation}}
\def\ee{\end{equation}}
\def\bea{\begin{eqnarray}}
\def\eea{\end{eqnarray}}

\newcommand{\Photo}{\includegraphics[height=35mm]{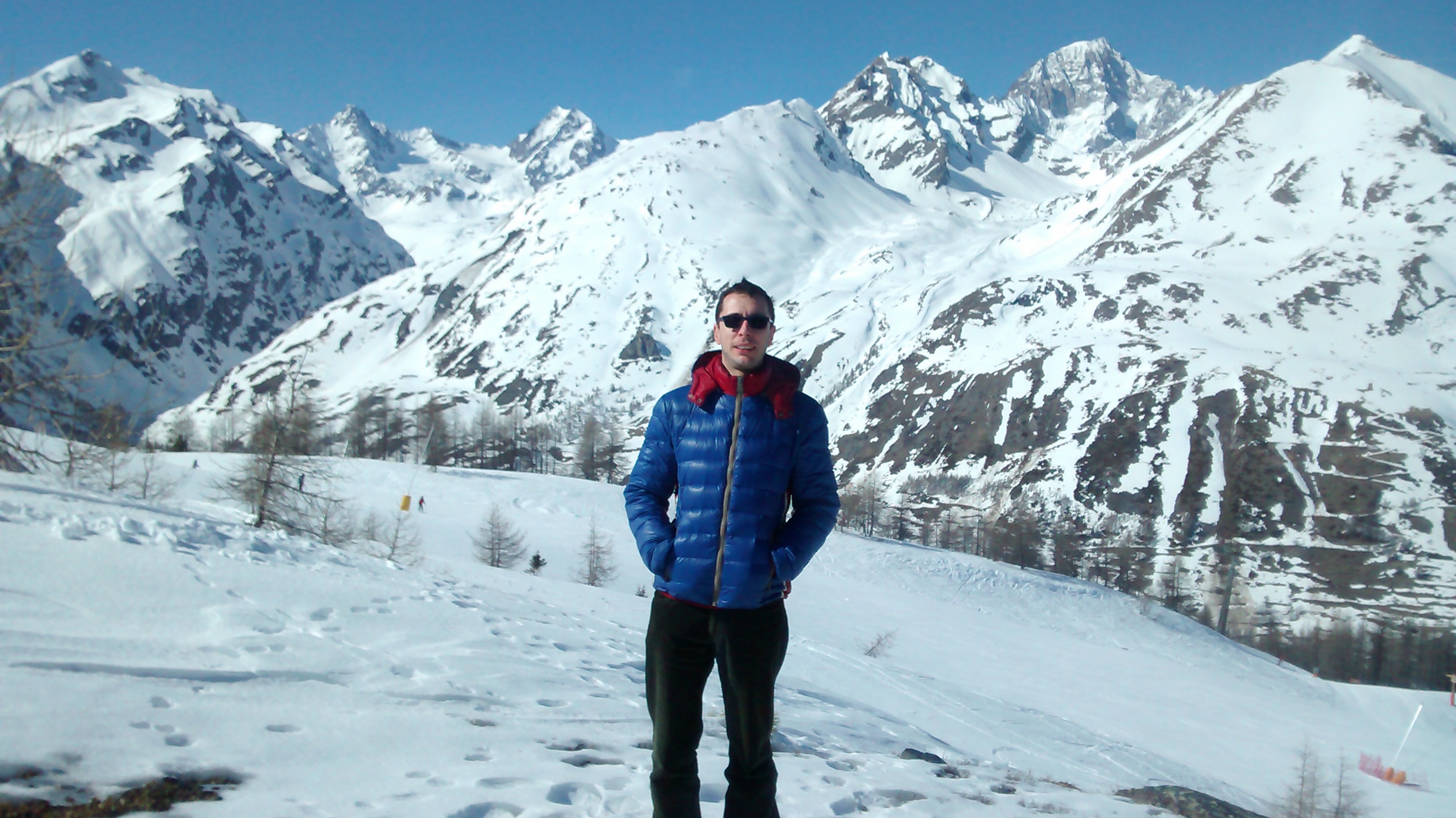}}

\begin{document}
\vspace*{4cm}
\title{THE MASS ACCRETION RATE OF GALAXY CLUSTERS: \\ A MEASURABLE QUANTITY}

\author{ C. DE BONI }

\address{Institut d'Astrophysique Spatiale, CNRS (UMR8617) Universit\'e Paris-Sud 11, B\^{a}t. 121, \\ F-91405 Orsay, France; \\ Dipartimento di Fisica, Universit\`a di Torino, via P. Giuria 1,
  I-10125 Torino, Italy; \\ Istituto Nazionale di Fisica Nucleare (INFN), Sezione di Torino, via P. Giuria 1,
  I-10125 Torino, Italy}

\maketitle\abstracts{
We~\cite{dsd} are interested in investigating the growth of structures at the nonlinear scales of galaxy clusters from an observational perspective:
we explore the possibility of measuring the mass accretion rate of galaxy
clusters from their mass profile beyond the virial radius.
We derive the accretion rate from the mass of a spherical shell whose infall
velocity is extracted from $N$-body simulations.
In the redshift range $z=[0,2]$, our prescription returns an average
 mass accretion rate within $20-40 \%$ of the average rate derived from the merger
trees of dark matter haloes extracted from $N$-body simulations.
Our result suggests that measuring the
mean mass accretion rate of a sample of galaxy clusters is actually feasible, thus providing a new potential observational test of the cosmological and structure formation models.}

\section{The Mass Accretion History}\label{sec:MAH}

The stochastic process of the mass accretion of dark matter halos is generally investigated with the identification of the merger trees of dark matter halos, enabling the study of the mass accretion history (MAH) and its derivative with respect to cosmic time, the mass accretion rate (MAR), as a function of redshift $z$.
Here we explore the possibility of using the cluster mass profile at radii larger than the virial radius to estimate the MAR of galaxy clusters by measuring the mass of an infalling spherical shell surrounding the cluster. 
We compare the MAR estimated with our recipe with the MAR of dark matter halos derived from their halo merger trees obtained from the CoDECS suite of numerical simulations (www.marcobaldi.it/CoDECS)~\cite{mb}.
For each halo at $z=0$ we build the MAH at $2R_{200}$. $2R_{200}$ roughly corresponds to the outermost radius reached by the accreted material in its first orbit around the cluster center, the so-called splashback radius $R_{\rm sp}$~\cite{mdk}.

\section{The Spherical Infall Model}\label{subsec:spherical}

We aim to quantify the mass accretion rate (MAR) as the ratio between the mass of a spherical shell of thickness $\delta_{s} R_{i}$ and the time it takes to fall onto the halo.
The thickness $\delta_{s}$ depends on the initial radius of the cluster $R_{i}$, on the infall time $t_{\rm inf}$, and on the initial infall velocity $v_i$.
The infall radius $R_{\rm inf}$, i.e., the radius where the minimum of the radial velocity profile 
$v_{\rm rad}$ occurs, is between $2R_{200}$ and $3R_{200}$, independently of mass and redshift (Figure~\ref{fig:figure}, left).
We use $R_i = 2R_{200} \sim R_{\rm sp}$ as the radius at which we consider the infall to happen in our spherical infall prescription and $v_i \sim v_{\rm rad}(R_{\rm inf})$.
For the infall time, which is the last parameter of the model, we choose $t_{\rm inf} = 10^9 \ {\rm{yr}}$, similar to the dynamical time for the clusters of our analysis.
Once $R_{i}$, $v_{\rm shell}$ and $t_{\rm inf}$ are specified, the model is completely determined. For each halo and for each progenitor at higher redshift, we evaluate the thickness $\delta_{s}$ of the infalling shell and its mass.

\section{Results}\label{sec:results}
 
\begin{figure}
\begin{minipage}{0.33\linewidth}
\centerline{\includegraphics[width=\linewidth]{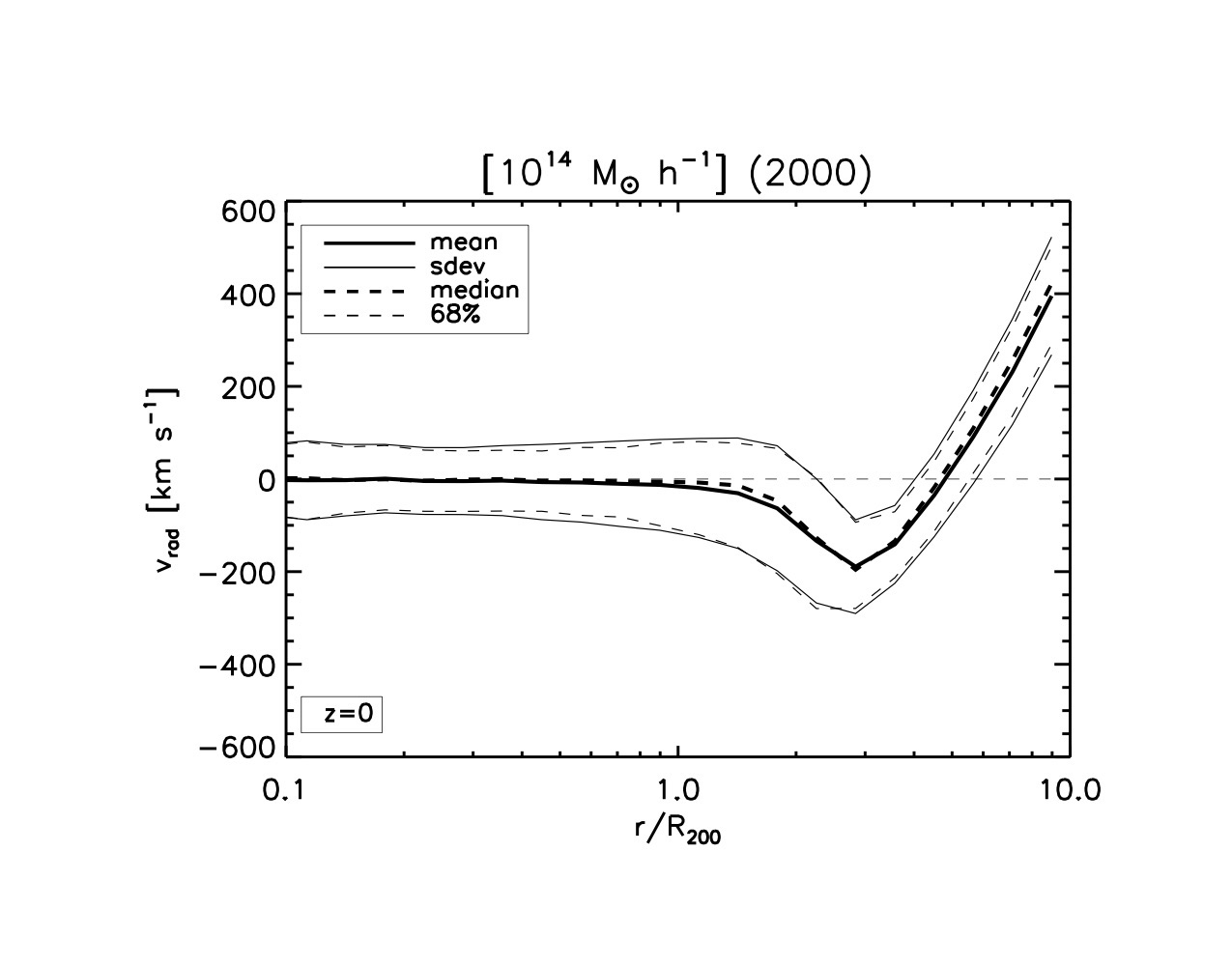}}
\end{minipage}
\hfill
\begin{minipage}{0.32\linewidth}
\centerline{\includegraphics[width=\linewidth]{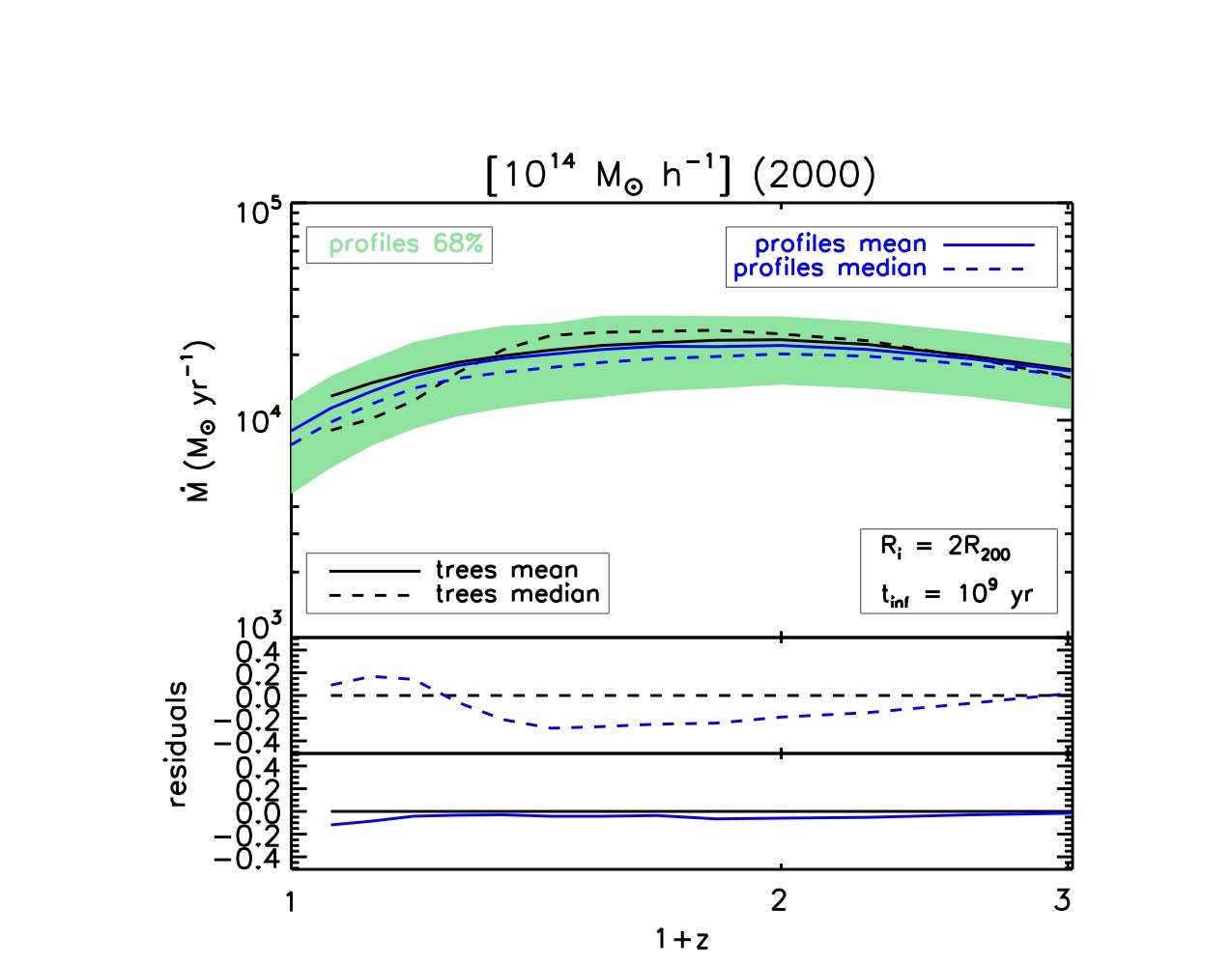}}
\end{minipage}
\hfill
\begin{minipage}{0.32\linewidth}
\centerline{\includegraphics[width=\linewidth]{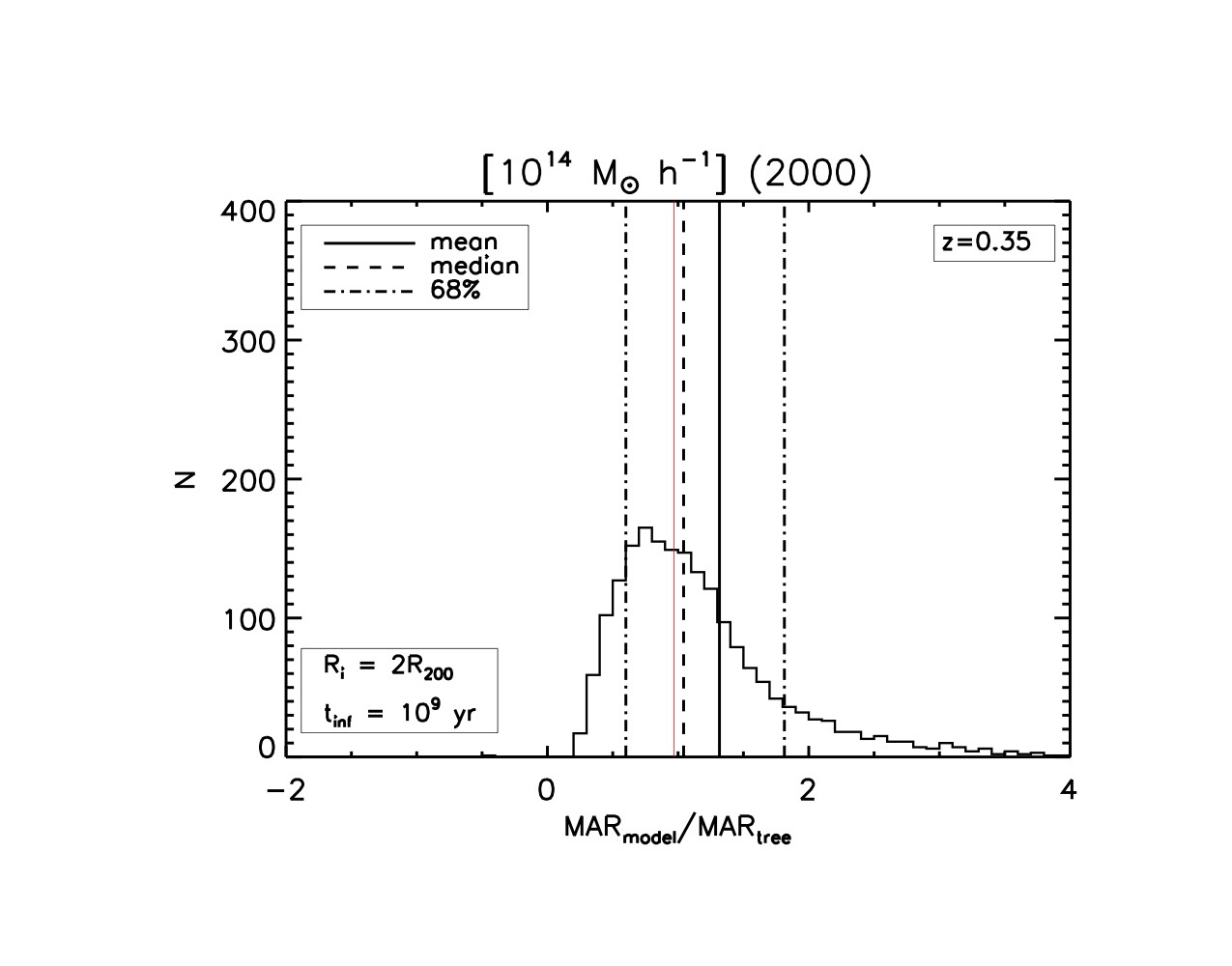}}
\end{minipage}
\caption[]{Radial velocity profile at $z=0$ (left), results of our spherical infall model and comparison with the MAR from merger trees (center) and histogram of the halo-by-halo ratio between the MAR from our spherical infall model and the MAR from the merger trees at $z=0.35$ (right) for clusters of $10^{14} \ {\rm{M_{\odot}}} \ h^{-1}$.}
\label{fig:figure}
\end{figure}

The mean and median results from the merger trees lie within the region defined by the 68\% range of the distribution of the MAR's obtained with our spherical infall prescription.
The mean and median MAR from our prescription recover the merger tree results within 20\% in the redshift range $z = [0,2]$ for objects of $10^{14} \ {\rm{M_{\odot}}} \ h^{-1}$ (Figure~\ref{fig:figure}, center).
Despite our recipe was not conceived to completely capture all the features of the MAR derived by the complex merging process of individual halos, the average of the MAR of individual halos still is satisfactorily estimated by our recipe.
The median value of the ratio between the MAR estimated with our recipe and the MAR derived from the merger tree for each individual halo is close to the ratio between the average MAR from our model and the average MAR from the merger trees (Figure~\ref{fig:figure}, right). The final goal is to apply our spherical infall recipe to the clusters in the CIRS and HeCS catalogs whose outer mass profiles have already been measured with the caustic technique.

\section*{Acknowledgments}

I thank my collaborators for allowing me to show results from our common projects.
This work was partially supported by grants from R\'egion Ile-de-France.

\end{document}